\documentclass[times, 11pt]{article}
\pdfoutput=1
\usepackage{graphicx}		   % for Figures
\usepackage{amssymb,amsmath}   % for mathbb and some special symbols
\usepackage{epstopdf}		   % for inclusion of eps figs in pdflaTeX
\usepackage{color,textcomp}    % for the Macintosh Symbols and URL coloring

 \renewcommand{\arraystretch}{1.2}

\newcommand{\Sec}[1]{\S \ref{sec:#1}}
\newcommand{\Fig}[1]{Fig.~\ref{fig:#1}}
\newcommand{\Tbl}[1]{Table~\ref{tbl:#1}}

\newcommand{\InsertFig}[4]
{\begin{figure}[ht]
       \centerline{
         \includegraphics[width=#4]{./figures/#1}
       }
       \caption{{\footnotesize  #2}
       \label{fig:#3}}
\end{figure}}

\newcommand{\InsertFigTwo}[5] {
\begin{figure}[htb]
       \centerline{
         \includegraphics[width=#5]{./figures/#1}
         \hskip 0.5in
         \includegraphics[width=#5]{./figures/#2}
       }
       \caption{{\footnotesize  #3}
       \label{fig:#4}}
\end{figure}}

\newcommand{\InsertFigFour}[7] {
\begin{figure}[ht]
       \centerline{ \renewcommand{\arraystretch}{0.01}
         \begin{tabular}{cc}
         \includegraphics[width=#7]{./figures/#1}&  \includegraphics[width=#7]{./figures/#2} \\
         \includegraphics[width=#7]{./figures/#3}  &  \includegraphics[width=#7]{./figures/#4}
        \end{tabular}
       }
       \caption{{\footnotesize  #5}
       \label{fig:#6}}
\end{figure}}

\newcommand{\R}{{\mathbb{ R}}}

\newcommand{\T}{{\mathbb{ T}}}
\newcommand{\Z}{{\mathbb{ Z}}}
\newcommand{\N}{{\mathbb{ N}}}

\newcommand{\Std}{\textit{StdMap}}

\newcommand{\tr}{\mathop{\rm tr}}
\newcommand{\floor}[1]{{\lfloor{#1}\rfloor}}

\newcommand{\Fix}[1] {\mbox{Fix}(#1)}

\newcommand{\beq}[1]{\begin{equation}\label{eq:#1}}
\newcommand{\eeq}{\end{equation}}

\newcommand{\cmd}{\raisebox{-0.4ex}{\includegraphics[width=.8em]{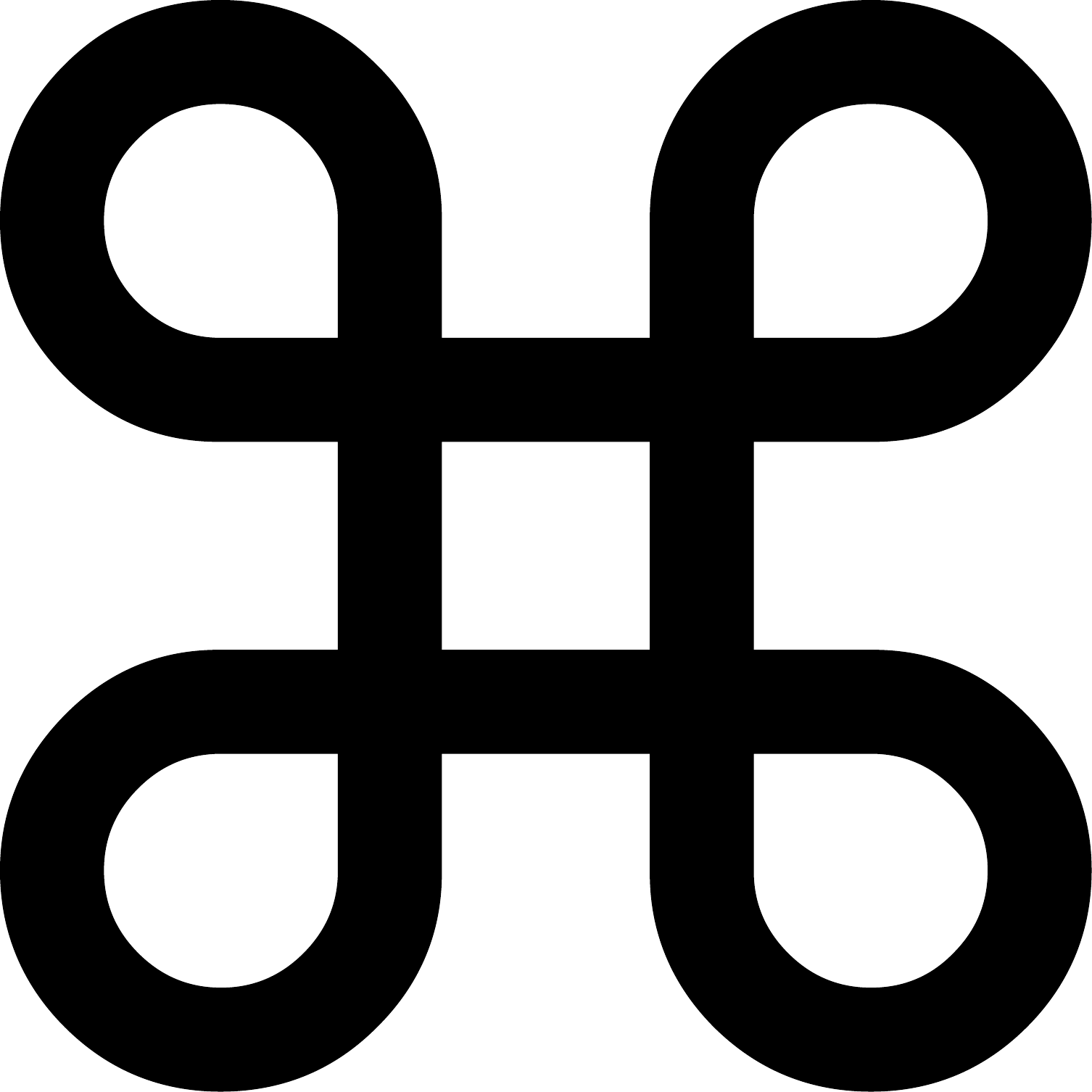}}}
\newcommand{\shift}{\raisebox{-0.12ex}{\includegraphics[width=.8em]{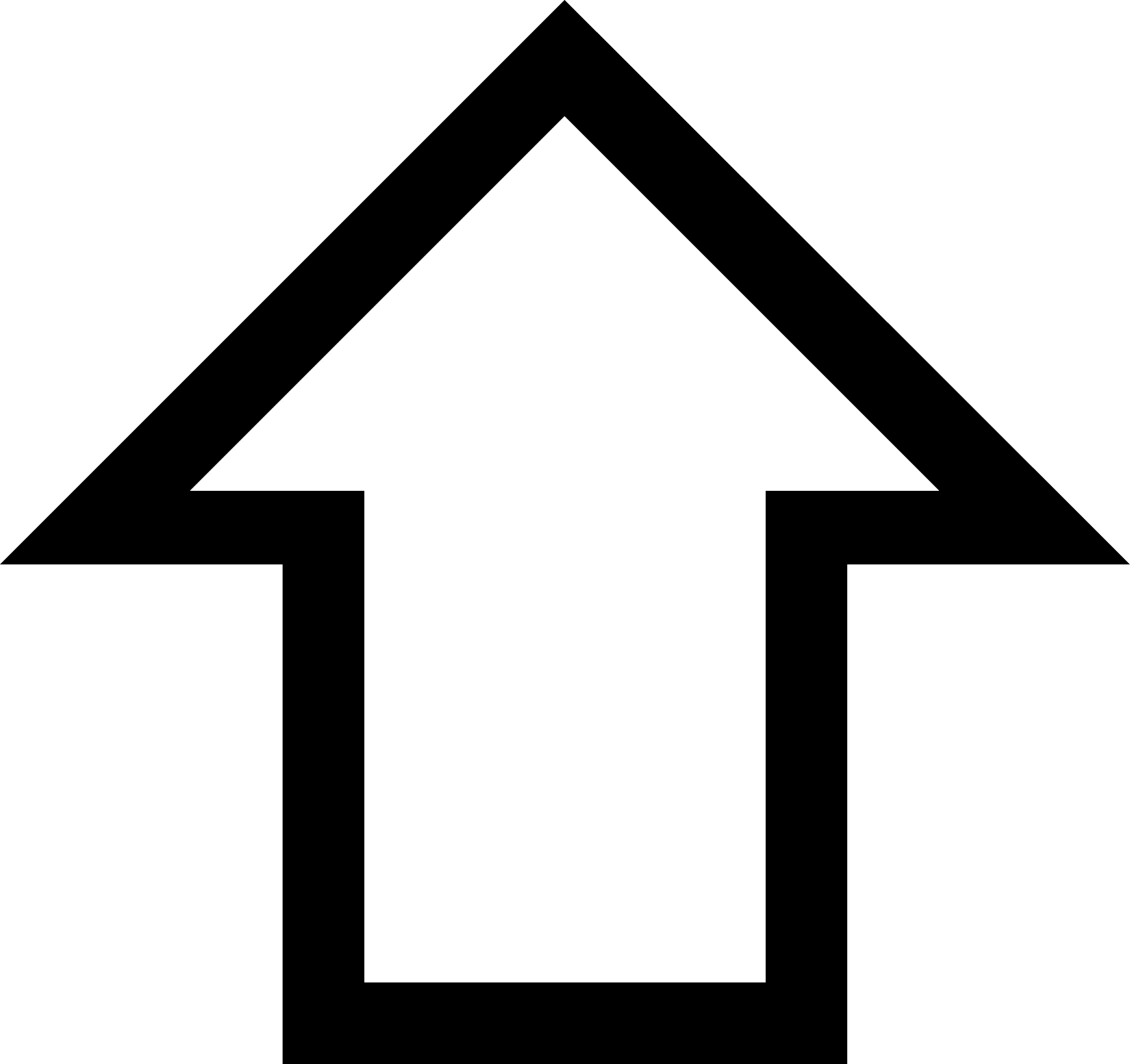}}}
\newcommand{\option}{\raisebox{-0.12ex}{\includegraphics[width=.8em]{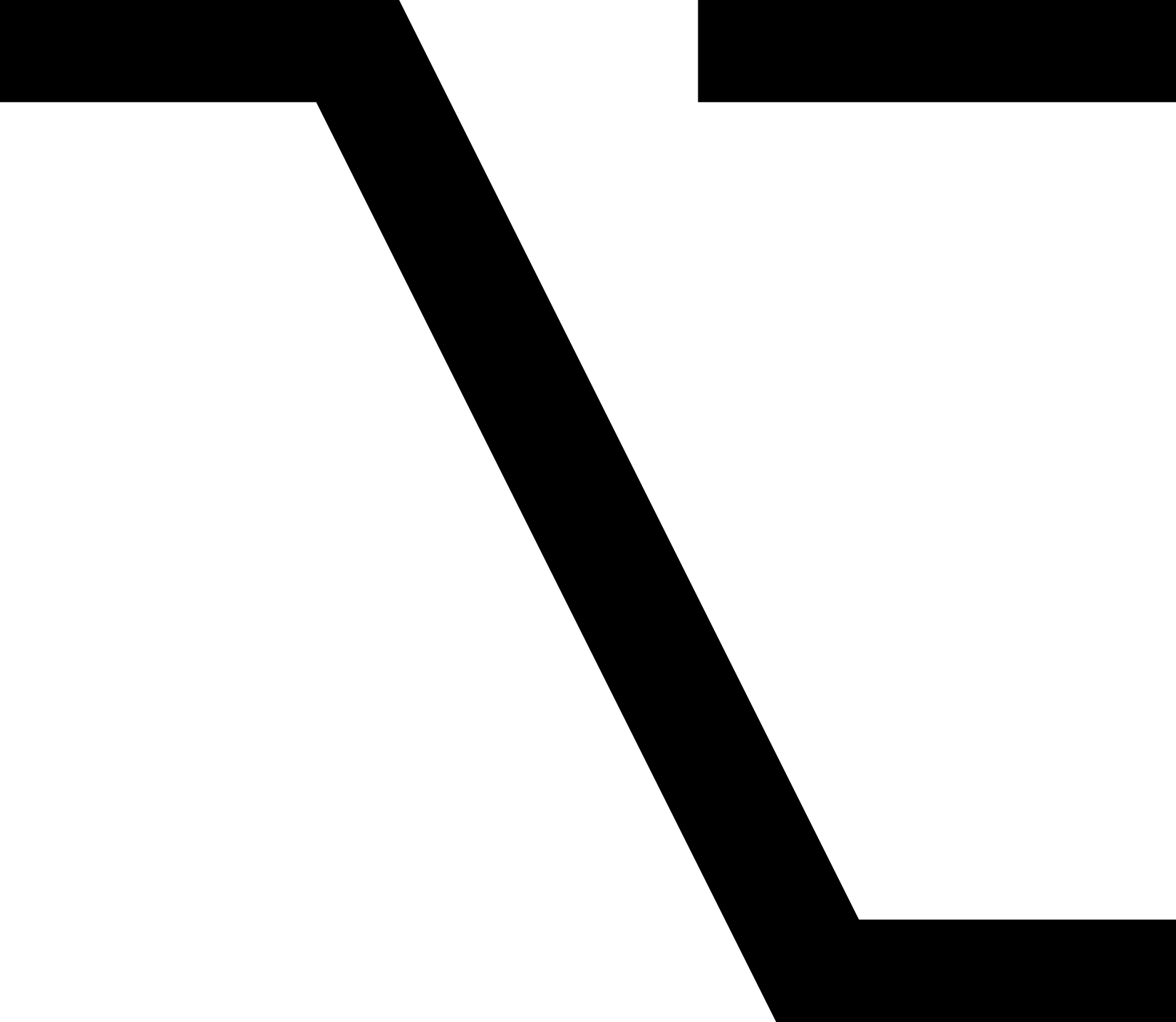}}}

\newcommand{\menu}[2]{\texttt{#1} $\to$ \texttt{#2}}
\newcommand{\mfont}[1]{\texttt{#1}}

\title{Visual Explorations of Dynamics: the Standard Map}
\author{  
        J.~D.~Meiss\thanks
      {
        James.Meiss@Colorado.EDU, 
        The author was supported in part by NSF grant DMS-0707659.
        A version of this paper was presented at the Perspectives in 
        Nonlinear Dynamics 2007 conference, supported by the Abdus Salam ICTP in Trieste.      
      } \\
     Department of Applied Mathematics\\
     University of Colorado \\
     Boulder, CO 80309-0526 \\
}
\date{\today}
\begin{document}
\maketitle
%\abstract{
\begin{abstract}
The Macintosh application \Std~allows easy exploration of many of the phenomena of area-preserving mappings. This tutorial explains some of these phenomena and presents a number of simple experiments centered on the use of this program. 

\begin{center}
	\textit{Dedicated to the memory of John Greene.} \\
	\textit{His ideas are as timeless as his generosity is legendary.}
\end{center}
\end{abstract}

%%%%%%%%%%%%%%%
%%  Introduction
%%%%%%%%%%%%%%%
\section{Introduction}

Area-preserving mappings give rise to incredibly rich dynamics and mathematics. They also arise in many applications including mechanics, chemistry, celestial dynamics, plasma physics, condensed matter physics, and many other fields.

The ``standard map" $z \to f(z) = z'$, where $z = (x,y) \in \T \times \R$, is defined by
\beq{stdmap}
   f: \begin{array}{lcl}
	      x' &=& x+ y -\frac{k}{2\pi} \sin(2\pi x) \; \mod 1 \;,\\
	      y' &=& y-\frac{k}{2\pi} \sin(2\pi x) \;.
	  \end{array}
\eeq
It was proposed by Boris Chirikov as a paradigm for resonance phenomena in conservative systems; it was also derived independently by Bryan Taylor as a model for the motion of charged particles in a strong magnetic field. The standard map is one of a class of maps called ``area-preserving twist maps;" it also has a number of symmetries, the most important of which is a time-reversal symmetry, see \Sec{reversible}.

The map \eqref{eq:stdmap} has a single parameter, $k$, that represents the strength of the nonlinearity. A number of physical systems can be modeled by this map. One is the cyclotron particle accelerator as described in my review article \cite{Meiss92}; another is the ``kicked rotor", described in \cite{Lichtenberg92}, and third is the Frenkel-Kontorova model of condensed matter physics studied extensively by \cite{Aubry83}. 

The application ``\Std" (available from $<$\textcolor{blue}{http://amath.colorado.edu/$\sim$jdm/stdmap.html}$>$) allows you to explore the dynamics of a dozen or so reversible, area-preserving mappings---including the standard map---on a Macintosh\texttrademark~computer.  It has an advantage over more general computer algebra tools of being compiled and dedicated to this one task (you can iterate hundreds of thousands of steps per second on any recent computer) and of being interactive. In this tutorial, I will describe some of the experiments that can be performed with \Std~and demonstrate some of the basic phenomena of conservative dynamical systems.

%%%%%%%%%%%
%%%% Starting
%%%%%%%%%%%
\section{Starting \Std}

\Std~is a Macintosh application and runs under Mac OS 10.2 or later. The most recent version is always available from $<$\textcolor{blue}{http://amath.colorado.edu/$\sim$jdm/stdmap.html}$>$; this tutorial will refer to version 4.5. The file on the web site is compressed; when you download it, it may automatically decompress, or you can double click on the ``.zip" file to create the executable application.\footnote{
%%%
	For the purposes of this introduction, if you have ever used \Std~before, you should delete the preference file $\sim$/Library/Preferences/edu.colorado.stdmap.plist, so that you can start the application in its default state.}
%%%

%%%%%
\InsertFig{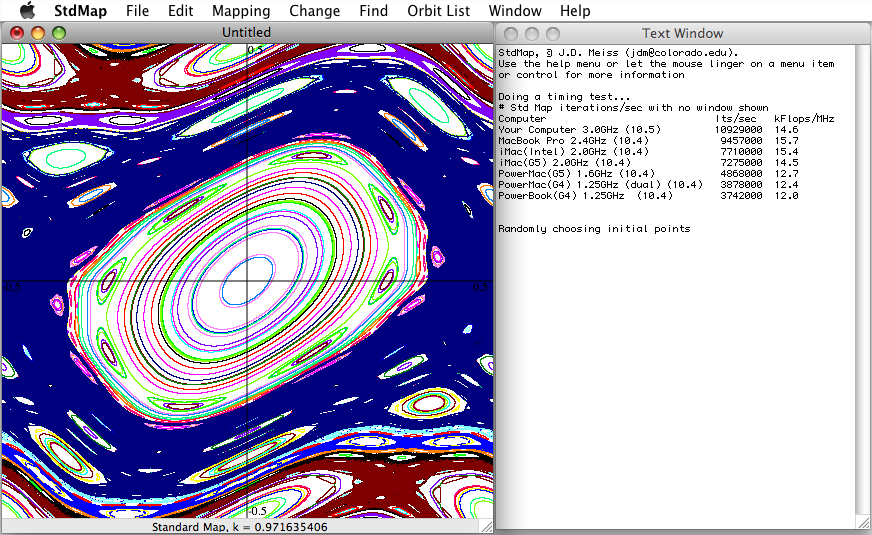}{Starting up the application \Std~ results in continuous iteration of \eqref{eq:stdmap} for randomly selected initial conditions. The text window provides information about the speed of your computer, which is used to select the global variable $N$, the number of iterates per function call.}{startup}{5in}
%%%%%

Double-click on the application to launch it. After a slight delay, two windows will open, as shown in \Fig{startup}; the left window contains a plot window and the right is a text window. By default, the program will start iterating \eqref{eq:stdmap} for $k = 0.971635406$ with random initial conditions.\footnote
%%%
{
	For the significance of this parameter value, see \Sec{greene}.
}
%%%

What you see on your screen, and in \Fig{startup}, are many orbits, each given a color from a list of 19 colors. Each orbit consists of a randomly selected initial point $z_0 = (x_0,y_0)$ and its next $N$ iterates, 
\[
(x_t,y_t) = f(x_{t-1} ,y_{t-1}) \;, \quad  1 \le t \le N \;,
\]
where $N$ is selected by the program depending upon the speed of your computer.\footnote
%%%%
{
	On my $3$ GHz machine, $N$ is 91074 to give an elapsed time for $N$ iterations 
	of about 1/60-second. You can view or change this using the \mfont{Change} menu 
	and selecting \mfont{Speedometer...}.
}
%%%%
Some of the orbits behave ``regularly" filling out topological circles or families of circles and some behave ``irregularly" or ``chaotically" filling out what seems to be a region of nonzero area. We will discuss this more in \Sec{chaos}.

The natural phase space of the standard map is the cylinder, $\T \times \R$, with the angle variable, $x$, having period-one, and the momentum, $y$, taking any real value. To implement this, the ``mod" operation for $x$  in \eqref{eq:stdmap} is defined in the program as 
\[
	x \mod 1 \equiv  x - \floor {x + 1/2}
\]
so that $x \mod 1 \in [-\frac12, \frac12)$. However, the standard map is actually periodic in the momentum direction as well, and the phase space can also be thought of as the two-dimensional torus if we add a mod $1$ operator to the $y$-equation; indeed, this is the default setting for \Std. The reason why this works is that the orbit starting at $(x_0, y_0)$ and the orbit starting at $(\xi_0,\eta_0) = (x_0 + m, y_0+n)$ for any $(m,n) \in \Z^2$ are related by
\begin{align*}
	\xi_t &= x_t  \\
	\eta_t  &= y_t + n
\end{align*}
thus the vertical separation by an integer stays constant, and $\eta_t \mod 1 = y_t$.
You can turn off the periodic modulus for $y$ by selecting the \mfont{Change} menu (see \Fig{menus}) and from the \mfont{Clipping} submenu, switching to \mfont{y unbounded}.

%%%%%%%%%%%%
%%%%% One Step
%%%%%%%%%%%%
\section{Iterating Step-by-Step}

Iteration of a specific initial condition in the standard map is easy: just click on the point in the plot window to start there. If you would like to start with a clean window, select \mfont{Clear} from the \mfont{Edit} menu or type \cmd-Y. Now click on a point near the origin and notice that it almost instantaneously fills in a small ellipse about the origin. This is because $(0,0)$ is a stable fixed point for this value of $k$, see \Sec{chaos}, and Moser's famous twist theorem, part of KAM theory, implies that the iterates near an elliptic fixed point will generically lie on invariant circles \cite{Meiss92, delaLlave01, Poshel01}.

You can also show the number of iterates on an orbit by selecting \menu{Change}{Show \# of Iterates}, and the position of the cursor in the window with \menu{Change}{Show Position} (\cmd-U), see \Fig{menus}. These settings will be remembered the next time you start \Std. 

Now click on an initial condition near the unstable fixed point at $(0.5, 0.0)$. This trajectory is chaotic and fills a large region of phase space; what is most surprising, though, is that this region is highly nonuniform, containing many holes. Numerical evidence indicates that the invariant set that is densely covered by a chaotic orbit is a \textit{fat fractal} (that is a fractal with positive measure), though this has never been proven \cite{Umberger86, Hanson87}. How long do you have to iterate until the trace of the orbit settles down and no new pixels are filled? If you have a monitor with many pixels, this time can be very long indeed \cite{Meiss94}. If we change the parameter value by selecting \menu{Change}{Map Parameters...}, and typing the value $2.0$ for $k$ in the \mfont{Parameter Dialog} window, the large chaotic region has the form shown in \Fig{fatfractal}.

%%%
\InsertFigTwo{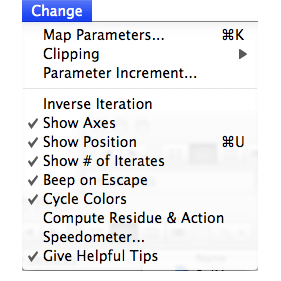}{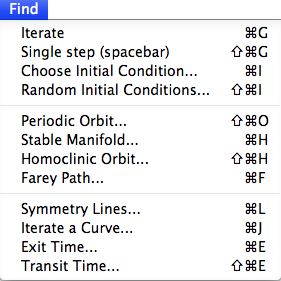}{The Change and Find menus for \Std. In this tutorial, we denote a menu selection using the $\to$ symbol, thus \menu{Change} {Show Axes} indicates that we are selecting the sixth item in the change menu. This item is checked since the axes are currently shown. Selecting it will toggle the display of the axes in the plot window. Your selections in this menu will be remembered the next time you start \Std.}{menus}{2.5in}
%%%

%%%%%
\InsertFigTwo{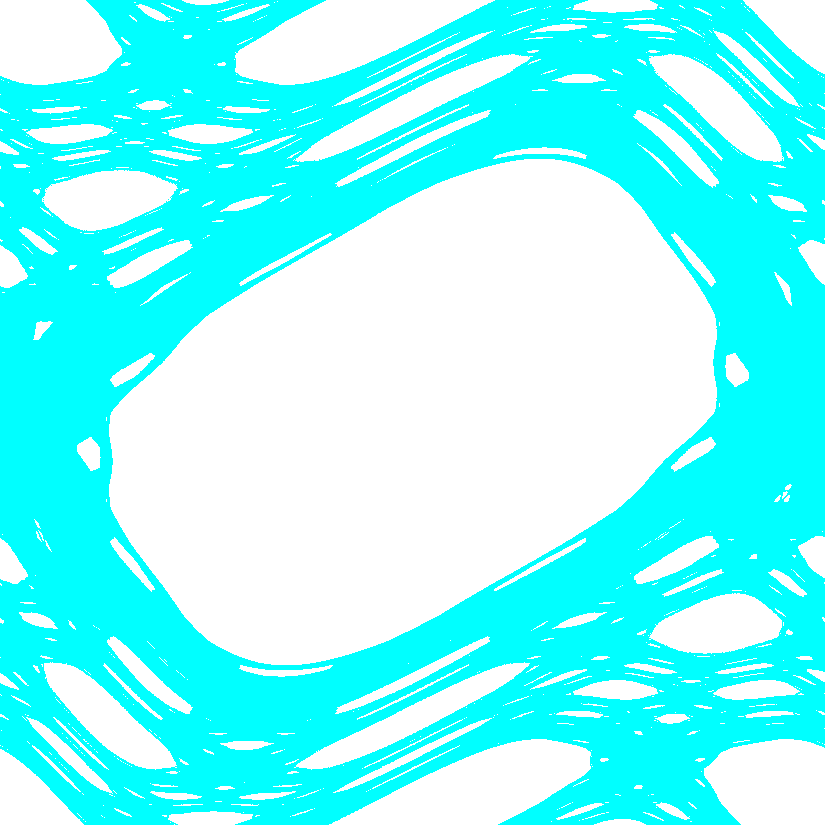}{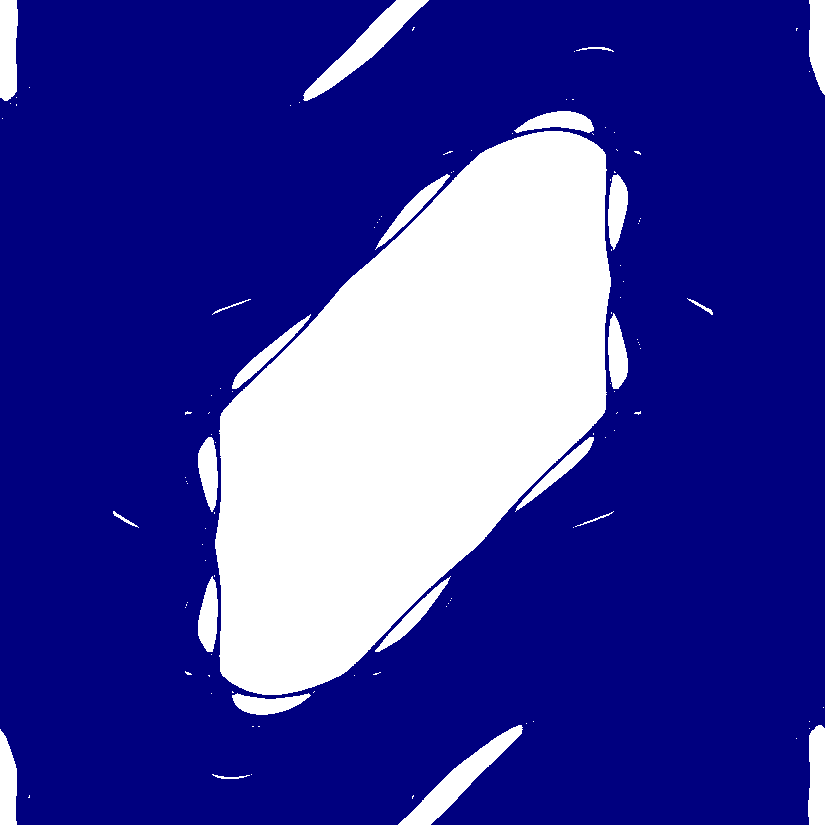}{Chaotic orbits appear to densely cover a fat fractal like this set generated by iteration of a single initial condition for $k = 1.0$ (left pane) and $k=2.0$ (right pane)}{fatfractal}{3in}
%%%%%

Often iteration in \Std~is too quick to really see what is going on; there are two ways to slow it down. One is to change the iteration mode in the \mfont{Find} menu. In particular \menu{Find}{Single Step (spacebar)} stops the iteration and moves the point forward (now drawn as a small square instead of a pixel) only when you hit the spacebar, recall \Fig{menus}. Let us also change the parameter of the map to something smaller so that there is less chaos, choose \menu{Change} {Map Parameters...}, and type the value $0.3$ for $k$ in the \mfont{Parameter Dialog} window. Note that the current value of $k$ is displayed at the bottom of the plot window.  Now when you click on an initial point, and repeatedly hit the spacebar to iterate step-by-step, you will see a portrait like that shown in \Fig{singleStep}.\footnote
%%%
{
	The second way to slow iteration is to decrease the global variable $N$ that represents
	the number of iterates done each function call when you are in continuous 
	iteration mode, \cmd-G. You can do this with \menu{Change}{Speedometer}. 
}
%%%

%%%
\InsertFig{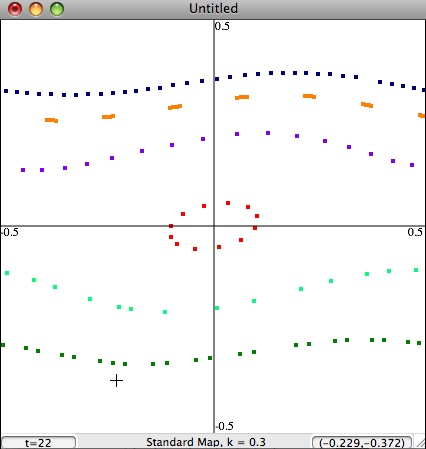}{Iterating the standard map one step at a time for $k = 0.3$.}{singleStep}{3in}
%%%

Iterating one step at a time reinforces the fact that maps are dynamical systems with discrete time. Another key feature of the map \eqref{eq:stdmap} is that the horizontal distance between successive iterations grows with the momentum value. Mathematically this is an example of the \textit{twist} condition, 
\beq{twist}
	\tau = \frac{\partial x'}{\partial y} \neq 0
\eeq
For the standard map, $\tau = 1$, and so it twists to the right. Perhaps a better way to visualize twist is to iterate a curve of initial conditions instead of a single point. You can do this in \Std~by selecting \menu{Find} {Curve...} or typing \cmd-J, recall \Fig{menus}. This will open the \mfont{curve dialog}, as shown in \Fig{curves}. There are five types of curves that you can iterate, and you can select one by clicking in one of the boxes. For this demonstration, click on the middle box, which selects the line type. Now click and drag a line in the plot window. A small dialog window will open near the bottom of the plot when you release the mouse. This shows the number of iterates that you have performed, either by clicking the \mfont{Iterate} button or hitting the spacebar or the I key. If you start with a vertical line,\footnote
%%%
{
   It is a standard Macintosh convention that drawing is \textit{constrained} if you 
   hold down the shift key. In this case, the line will be constrained to be exactly 
   vertical, horizontal or at forty-five degrees. Constrained drawing is also an easy way
   to draw squares and circles. 
}
%%%
as we did in \Fig{curves}, the first iterate will be the line $y = x$---the vertical line twists to the right and becomes a graph over $x$. This is the geometrical meaning of \eqref{eq:twist}. The second iterate of this line becomes the curve
\[
	(x,y) = (2s-\frac{k}{2\pi}\sin(2\pi s), s-\frac{k}{2\pi}\sin(2\pi s))
\]
which is still a graph over $x$ for $|k| < 2.0$. When $k = 0.3$, the curve is no longer a graph  on the sixth iterate---it is triple-valued for some intervals of $x$, see \Fig{curves}. Nevertheless, this curve still has some aspects of twist: it \textit{tilts} to the right. The map $f^n$ is a composition of twist maps, and is called a \textit{tilt map} \cite{Boyland88}. Tilt maps have many of the properties of twist maps, and it has even been recently shown that they have variational principles, so that KAM and Aubry-Mather theory can be applied \cite{Hu98}.

%%%
\InsertFigTwo{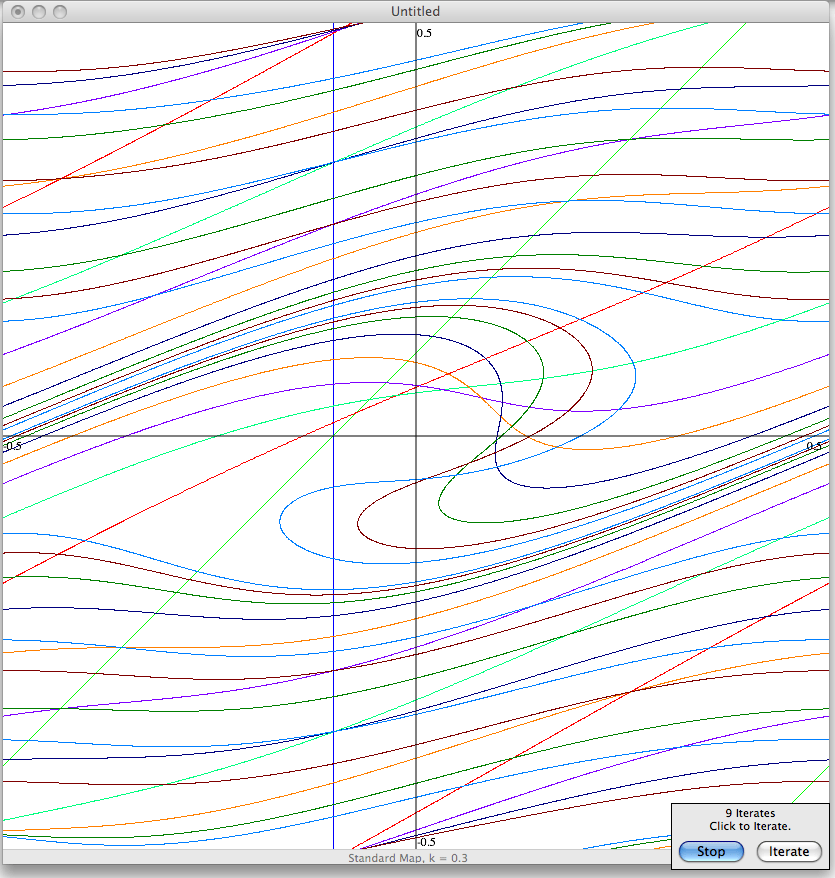}{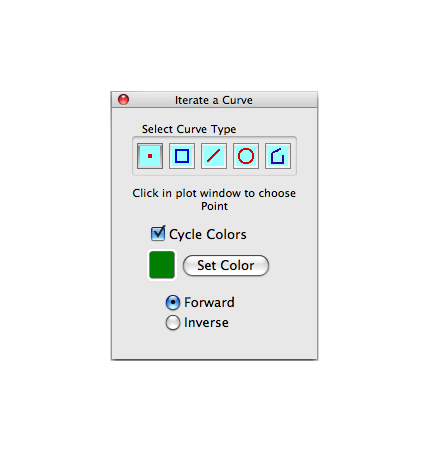}{You can iterate a curve by opening the curve dialog from the \mfont{Find} menu and clicking on one of the five curve types. Click and drag in the plot window to create the curve, then click on the \mfont{Iterate} button for each step, and finally \mfont{Stop} to finish. The five curve types are (1) small boxes, (2) rectangles, (3) lines, (4) polygons and (5) piecewise linear curves. In the last case each click forms a vertex; to make the last vertex, double-click. The left pane shows $9$ iterates of the vertical line $x = -0.1$ when $k = 0.3$.}{curves}{3in}
%%%

%%%%%%%%%%%%
%%%%% Chaos
%%%%%%%%%%%%
\section{The Onset of Chaos}\label{sec:chaos}

The standard map is \textit{integrable} when $k = 0$. Indeed for this value of $k$, the momentum is an invariant, and all the orbits lie on horizontal curves. More generally, an area-preserving map is integrable \cite{Veselov91, Meiss92} if there exists a non-constant function $I(x,y)$ such that
\[
	I \circ f = I \;.
\]
There are several other integrable maps that can be studied in \Std; for example, the McMillan map,
\beq{mcmillan}
	(x',y') = \left(x,-y - ax -b -\frac{1}{x} \right) \;,
\eeq
is integrable whenever $a = 1$ or $a = 0$, for any $b$ \cite{McMillan71}. The second is an
example of a \textit{generalized} standard map,
\beq{genStd}
	(x',y') = (x+ y+F(x), y+F(x)) \;,
\eeq
where the forcing function is 
\beq{suris}
	F(x) = -\frac{2}{\pi} \arctan\left( \frac{a \sin(2\pi x)}{1+a\cos(2\pi x)} \right) \;.
\eeq
This map and generalizations of it were proved to be integrable by Suris \cite{Suris89}.  Note that \eqref{eq:suris} limits on the standard map as $a \to 0$, indeed when $|a|< 1$,  $F$ has a simple Fourier series:
\[
	F(x) = \frac{2}{\pi} \sum_{n=1}^{\infty} \frac{(-a)^n}{n} \sin(2\pi n x)
\]
whose first term is just the standard map with $k = 4a$. The invariant functions for these maps are more complicated than that for the standard map; however, in each case there exists an analytic invariant $I(x,y)$. You can select either of these maps and study their dynamics using the \mfont{Mapping} menu.\footnote
%%%
{
	In \Std, the Suris map has a second term $\pi b \sin(2\pi x)$ that 
	destroys integrability when $b \neq 0$. Type \option-\cmd-1 to access it; 
	here \option~is the option key---hold this 
	key down and then type \cmd-1. This map was studied in \cite{Lomeli00}.
}
%%%

Now return to the standard map at $k=0$. Since the phase space is periodic in $x$, the orbits lie on topological circles, $y = c$; we call them \textit{rotational circles} since they encircle the cylinder, as opposed the \textit{librational} circles that lie near the origin    for $k > 0$, recall \Fig{startup}. What happens to the rotational circles as $k$ increases? This is the subject of KAM theory---the theory proposed by Kolmogorov and developed by Arnold and Moser. A nice way to visualize this with \Std~is to start a trajectory and increment the parameter slowly.

First set the parameter $k$ to $0$ by typing \cmd-k and entering $0$ in the dialog. Now click in the plot window to start an orbit (if you are still in single-step mode, type \cmd-G to start continuous iteration). The up-arrow and down-arrow keys can be used to increment $k$. The step size, $\Delta k$, is set using \menu{Change}{Parameter Increment...}; by default it is set to $0.05$. Repeatedly hitting the up-arrow key will restart the orbit at the new parameter value using the current point as the initial condition. You should see a sequence of pictures like those overlaid in \Fig{increments}.

In the left pane of \Fig{increments}, the initial orbit is selected with a small value of $y$; when $k = 0.1$, it is deformed to a rotational invariant circle, but when $k = 0.2$, it is trapped into the main \textit{resonance} about the fixed point at $(0,0)$. This librational invariant circle evolves slowly until $k = 2.25$, when the orbit is trapped in a chaotic layer about a period-$4$ saddle. In the right pane, the orbit is started with a larger $y$ value, and it remains a rotational invariant circle until $k = 0.85$, when it is trapped in a chaotic layer surrounding the main resonance. Note that the steps that you will see in your experiment depend in detail on the initial condition, the step size, and the current position $(x_t,y_t)$ when $k$ is incremented. Nevertheless, the phase space of the standard map consists primarily of regular invariant circles (some rotational and some librational) when $k < 0.5$ and primarily of chaotic orbits when $k > 3$.

It is also possible to increment $k$ with the up-arrow key while you are in random iteration mode, \menu{Find}{Random Initial Conditions...}, or using \shift-\cmd-I, where the symbol \shift~represents the shift key in the menus, recall \Fig{menus}

%%%%%
\InsertFigTwo{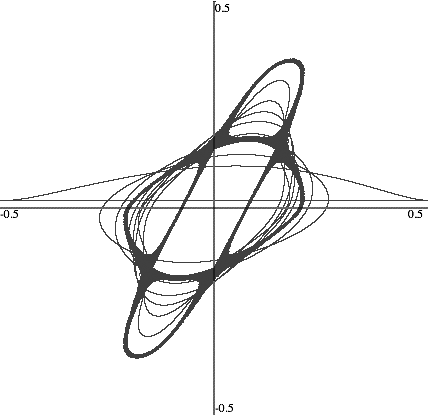}{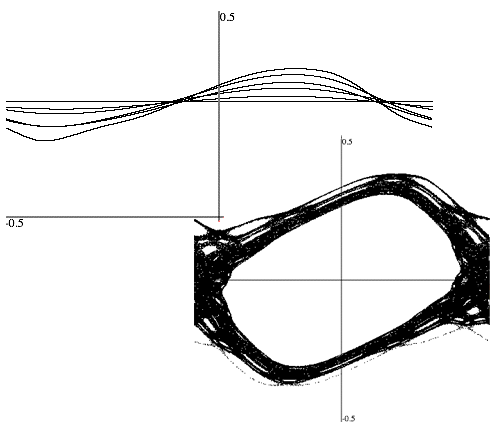}{Overlay showing the evolution of two orbits beginning at $k = 0$ as $k$ is incremented. In the left pane, an initial circle $y \approx 0.02$ evolves to the chaotic trajectory near a period-$4$ saddle when $k = 2.25$. In the right pane, the initial circle at $y \approx 0.3$ evolves to chaotic orbit shown in the inset at $k = 0.85$.}{increments}{3in}
%%%%%

Where does the chaos come from? A partial answer is: through bifurcations of periodic orbits. An orbit $\{(x_t,y_t): t \in \Z\}$ is periodic if there is a $q \in \N$ such that $(x_q, y_q) = (x_0,y_0)$; the least $q$ for which this is true is the \textit{period} of the orbit. For dynamical systems on the cylinder it is convenient to partially classify periodic orbits by their rotation number, $\omega = \frac{p}{q}$, were $p$ is the number of times the orbit encircles the cylinder. This can be easily computed by omitting the mod 1 in \eqref{eq:stdmap}, so that $x \in \R$; the resulting map is a \textit{lift} of the standard map to $\R^2$. For the lift, an orbit is $(p,q)$-periodic if $(x_q,y_q) = (x_0+p, y_0)$.

Periodic orbits can be computed in \Std~by specifying the pair $(p,q) \in \N^2$. To do this select \menu{Find}{Periodic Orbit...}. For example selecting $(p,q) = (2,5)$ results in the output 
\begin{equation}\label{eq:twofifth}
	(2,5) \mbox{ at } (-0.5, 0.262766939830826)
\end{equation}
when $k = 2.0$. This particular period-5 orbit is a saddle.\footnote
%%%
{
	The \mfont{initial condition} dialog has a pop-up menu labeled \mfont{symmetry}. The default setting for this is \mfont{minimize}, and this gives the orbit above. Choosing \mfont{minimax} from this menu will give the elliptic $(2,5)$ orbit. The symmetries of periodic orbits will be discussed in \Sec{reversible}.
}	
%%%
We can determine this by turning on the calculation of the \emph{residue} from the \mfont{Change} menu. John Greene introduced the residue as a convenient stability index for area-preserving maps in 1968 \cite{Greene68}. Let $Df$ denote the Jacobian matrix of the map; for \eqref{eq:stdmap},
\beq{Df}
	Df(x,y) = \begin{pmatrix}  1-k\cos(2\pi x) & 1 \\
								-k\cos(2\pi x) & 1 
			  \end{pmatrix} \;.
\eeq
The linearization of $f$ about a period-$n$ orbit $\{z_0,z_1,\ldots,z_{n-1}\}$ is the matrix
\[
	M = \prod_{t=0}^{n-1}  Df(z_t) = Df(z_{n-1}) Df(z_{n-2}) \ldots Df(z_1) Df(z_0)\;,
\]
and the eigenvalues $\lambda_\pm$ of $M$ are the \textit{multipliers} that determine the linear stability of the orbit. Since $f$ is area-preserving, $\det Df =1$, so that $\det M = \lambda_+ \lambda_- = 1$. Thus the eigenvalues of $M$ are completely determined by its trace, $\tr M = \lambda_+ + \lambda_-$. Greene defined the residue as
\beq{residue}
	R = \frac14 ( 2 -\tr M) \;.
\eeq
Since the product of the two multipliers is $1$, it is easy to see that orbits come in four classes:
\begin{enumerate}
	\item hyperbolic saddle when $R < 0$ with multipliers $0 < \lambda_- < 1 < \lambda_+$;
	\item elliptic when $ 0 < R < 1$ with multipliers $\lambda_{\pm} = e^{\pm2\pi i \omega}$ 
	on the unit circle;
	\item reflection-hyperbolic saddle when $R > 1$ with multipliers $ \lambda_+ < -1 < \lambda_- < 0$; and
	\item parabolic when $R = 0$ or $1$ with a pair of multipliers $\lambda_+ = \lambda_- = \pm 1$.
\end{enumerate}

In particular \eqref{eq:stdmap} has two fixed points, the origin and $(0.5,0) \equiv (-0.5,0)$. From \eqref{eq:Df}, the residue of a fixed point is simply $R = \frac{k}{4} \cos(2\pi x)$, and so origin has residue
\[
	R_{(0,0)} = \frac{k}4 \;,
\]
implying that it is elliptic when $0 < k < 4$ and reflection-hyperbolic for $k > 4$. This can easily be seen in \Std~by incrementing $k$ (using the up-arrow and down-arrow keys) and studying the orbits near the origin: when $k$ reaches $4$ the origin undergoes a period-doubling bifurcation and becomes unstable. Indeed one can show that there is an infinite sequence of such period doublings that branch from this orbit; they accumulate in a universal way that was first discovered by Feigenbaum for dissipative maps and by Greene and collaborators for the area-preserving case \cite{Greene81}.

By contrast, the residue of the second fixed point is
\[
  R_{(0.5,0)} = -\frac{k}{4} \; ;
\]
this always negative for $k > 0$, so that this point is a saddle with positive multipliers.

Turning on the residue calculation for the $(2,5)$ orbit at $k=2.0$ and finding the orbit again, gives $R = -11.88$, confirming that this orbit is also a regular saddle. Indeed, if you start iterating at this orbit, by selecting \menu{Find}{Continuously} (which automatically fills in the initial condition of the last orbit) you will find that the instability leads to numerical error: the iteration falls off the periodic orbit and rapidly covers the chaotic fat fractal shown in \Fig{fatfractal}. Another way to see this is to find the $(2,5)$ saddle, \menu{Find}{Periodic Orbit...}, and then immediately select \menu{Find}{Single Step} to start single-step iteration at the point \eqref{eq:twofifth}. As you iterate with the spacebar, you will see that the orbit visibly deviates from the periodic orbit after about 40 steps. For the given residue, the unstable eigenvalue of $M$ is $\lambda_+ = 49.5$, so that any error grows by a factor of $\lambda_+^{N/5} \approx 4 \times 10^{13}$ by $N = 40$; thus given the inevitable truncation error of double-precision floating point computations, it is not surprising that the orbit is lost.

There are better ways of investigating the properties of hyperbolic saddles, as we describe next.

%%%%%%%%%%%%
%%%%% Stable
%%%%%%%%%%%%
\section{Stable Manifolds}\label{sec:stable}
The stable and unstable sets, $W^s$ and $W^u$, of an invariant set $\Lambda$ are defined as\beq{basins}\begin{split}
	W^s_\Lambda &\equiv \{(x,y): f^t(x,y) \to \Lambda \mbox{ as } t \to \infty \} \;, \\
	W^u_\Lambda &\equiv \{(x,y): f^t(x,y) \to \Lambda \mbox{ as } t \to -\infty \} \;.
\end{split}\eeq
If $\Lambda$ is hyperbolic, then these sets are smooth submanifolds that are tangent to the eigenspace of the linearization of the map at $\Lambda$ \cite{Robinson99, Meiss07a}. 
For example, at the hyperbolic fixed point $(0.5,0)$, the linearization \eqref{eq:Df} has eigenvectors
\[
	v_\pm = \begin{pmatrix} 1 \\ 1-\lambda_\mp \end{pmatrix}
\]
with $\lambda_\pm = \frac12( 2+k \pm \sqrt{k(k+4)})$. When $k > 0$ the $+$-eigenvector corresponds to the unstable direction, and since $\lambda_- < 1$, it has a positive slope. The stable direction has negative slope. The stable manifold theorem implies that $W^{u,s}$ are smooth curves that start at $(0.5,0)$ with slopes $1-\lambda_\mp$.

To find these, \Std~uses an algorithm suggested by Dana Hobson \cite{Hobson93}. Selecting the command \menu{Find}{Stable Manifold...} (\cmd-H) opens a dialog with three text fields.\footnote
%%%
{
	There is also a pop-up menu for selecting the symmetry of the orbit. Leave this set
	at the default \mfont{minimizing} as this corresponds to the hyperbolic saddle.
}
%%%
The first two, $p$ and $q$, define the rotation number of the periodic orbit that will be the invariant set $\Lambda$. The third field is a sign, $s = \pm 1$, that determines which branch of the manifold to follow, i.e., one of the directions $\pm v$ for eigenvector $v$. The default values $(p,q) = (0,1)$ and $s = 1$ will plot the upward-going unstable manifold for the hyperbolic fixed point. After entering the three values, the program first finds a point $z^*$ on the $(p,q)$ orbit and computes its eigenvectors, then it allocates a large block of memory to store the computed points,\footnote
%%%
{
	We allocate an array that is no more than half the physical memory in your computer and 
	no larger than needed to contain up to 10 times the number of pixels in the plot window.
}
%%%
and finally it selects an initial point $z_0 = z^* + s \epsilon v_+$  where $\epsilon$ is chosen so that the distance $\lambda_+ \epsilon |v_+|$ is about one pixel. Clicking on the \mfont{Iterate} button or hitting spacebar iterates this point $q$ steps. The two points $\{z_0, z_q\}$ form an approximation to a \textit{fundamental segment} of $W^u$ for the $(p,q)$ orbit. If these points are more than one pixel apart the program inserts a new point between them using linear interpolation.

Hitting spacebar again causes all of the points on the fundamental segment to be iterated $q$ more times, and more points to be interpolated as needed. The result is a growing curve forming the unstable manifold. The color scheme follows a convention suggested by Bob Easton: red---for unstable---represents the blood moving away from the heart, and blue---for stable---represents the blood returning. The fundamental segment is stored in the large array and iteration will stop if this array fills up. Examples of unstable manifolds for $k = 1.0$ and $2.0$ are shown in \Fig{stable}. Though $W^u$ begins life as a nearly straight line along $v_+$, it rather quickly develops intricate whorls and tangles, and to the approximation of the computer screen appears to densely cover regions of phase space.

You will notice that stable manifolds (blue curves) are also computed by \Std---these are computed essentially for free using the reversing symmetry, see \Sec{reversible}. What is especially interesting is that the stable and unstable manifolds do not coincide, and that when they intersect, they usually cross at a nonzero angle; indeed for the $(0,1)$ orbit this was proven by \cite{Laz89, Gel01}. Points that are on both the stable and unstable manifolds of an invariant set $\Lambda$, were termed \textit{homoclinic} points by Poincar\'e, who discovered them. By the definition \eqref{eq:basins}, homoclinic points are both forward and backward asymptotic to $\Lambda$. The complexities of this picture were first envisioned by Poincar\'e who famously said:

\begin{quote}
When one tries to depict the figure formed by these two curves and their infinity of intersections, each of which corresponds to a doubly asymptotic solution, these intersections form a kind of net, web or infinitely tight mesh; neither of the two curves can ever cross itself, but must fold back on itself in a very complex way in order to cross the links of the web infinitely many times. One is struck by the complexity of this figure that I am not even attempting to draw \cite{Poincare1892}.
\end{quote}

You should immediately notice that there is a strong correlation between the chaotic fat fractal and the set covered by the invariant manifolds: compare \Fig{stable} with \Fig{fatfractal}. Indeed, one reasonable hypothesis is that the closure of $W^u_{(0.5,0)}$ is identical to the connected chaotic, fat fractal containing the saddle fixed point. The transverse intersection of stable and unstable manifolds gives rise to the famous horseshoe structure discovered by Stephen Smale in the 1960's \cite{Smale98}, and the existence of a transverse homoclinic orbit is one of the key features of chaos. As an exercise, you can also plot the stable and unstable manifolds for the Suris map (\option-\cmd-1) for the integrable case, $b = 0$, to see that its stable and unstable manifolds coincide.\footnote
%%%
{
	Do not set $a = 1$ for the Suris map, as in this case $F(x) = 2x$ and every point is parabolic and has period-4. The Suris map approximates the standard map with $a \approx k/4$, so try $a = \frac14$ to obtain a picture similar to the left pane of \Fig{stable}.
}
%%%

%%%%%
\InsertFigTwo{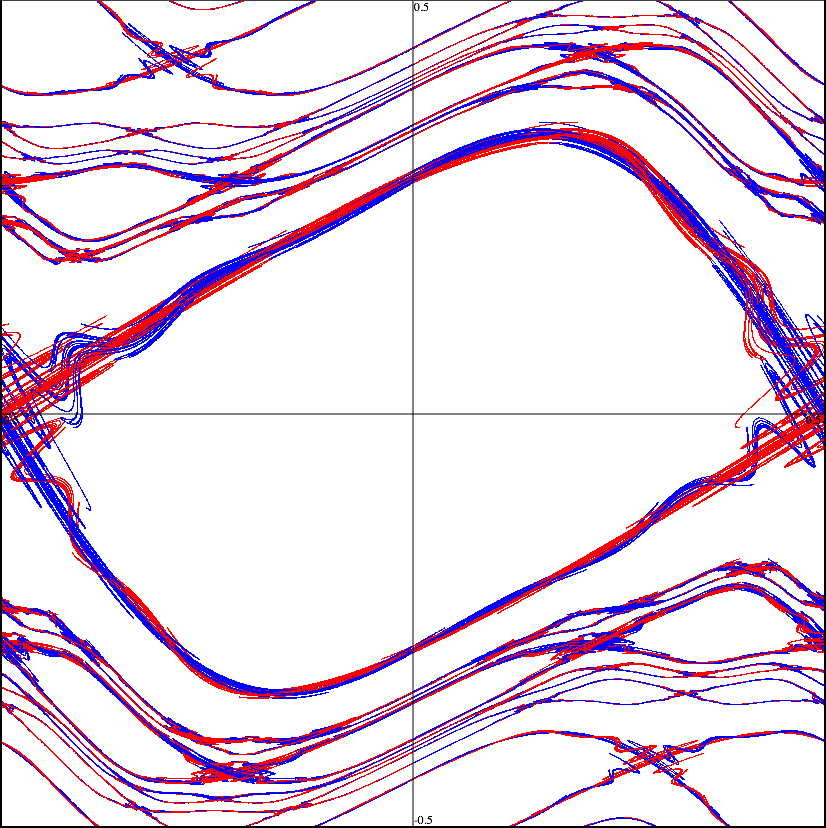}{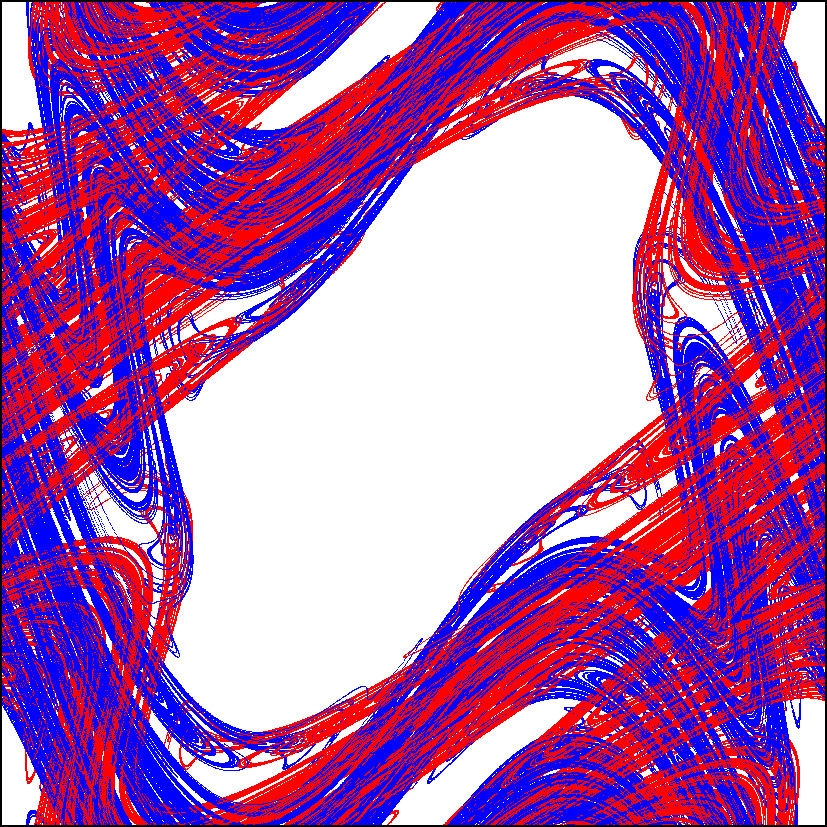}{Stable and unstable manifolds for the standard map at $k=1.0$ (left pane) and $k = 2.0$ (right pane). The left pane shows manifolds for the orbits $(1,2)$, 
$(2,5)$, $(3,8)$, $(1,3)$, $(2,7)$, $(0,1)$, $(-2,7)$, $(-1,3)$, $(-3,8)$, $(-2,5)$, $(-1,2)$, The right pane shows only the manifolds for $(0,1)$.}{stable}{3in}
%%%%%

%%%%%%%%%%%%
%%%%% Reversible
%%%%%%%%%%%%
\section{Reversing Symmetries}\label{sec:reversible}
All of the maps in \Std~are \textit{reversible}: there is an diffeomorphism $S$ that reverses the map
\beq{reversor}
	 f \circ S = S \circ f^{-1} \;.
\eeq
In this section we will explain why reversible maps are easier to study numerically than their irreversible cousins. 

Inversion of a map in \Std~is done using \eqref{eq:reversor} rather than writing a separate subroutine for $f^{-1}$. You can iterate backwards in \Std~by selecting \menu{Change}{Inverse Iteration}. The inverse will not look very different unless you are doing single steps, using \menu{Find}{Single Step}, \shift-\cmd-G, or iterating curves using \menu{Find}{Curve...}, \cmd-J.

For the generalized standard map \eqref{eq:genStd} with an odd force $F(-x) = -F(x)$, one choice for a reversing symmetry $S$ is 
\[
	S_1(x,y) = (-x, y + F(x)) \;.
\]
Indeed, it is easy to check that the inverse of \eqref{eq:genStd} is 
\[
	f^{-1}(x,y) = S_1 \circ f \circ S_1 = (x-y, y-F(x-y)) \;.
\]

Note that $S_1 \circ S_1 = id$, so that $S_1$ is an \textit{involution}. A consequence is that
\[
	S_2 = f \circ S_1(x,y) = (y-x, y)
\]
is also a reversor for $f$, and is an involution as well. Moreover these reversors are orientation-reversing, since their Jacobians have a negative determinant. Thus the generalized standard map has been factored into a composition of two orientation-reversing involutions, $f = S_2 \circ S_1$.

One of the significant implications of reversibility is that there are many symmetric periodic orbits, namely orbits invariant under $S$. It is not hard to see that any symmetric orbit must have points on the \textit{fixed} set $\Fix{S} \equiv \{z: S(z) = z \}$. For the reversors of the generalized standard map these are
\beq{dominant}\begin{split}
	\Fix{S_1} &= \{ (0,y)\} \;, \\
	\Fix{S_2} &= \{ (x,y) : x = y\} \;.
\end{split}\eeq

Maps can also have other symmetries that conjugate the map to itself, or \textit{commute} with the map:
\[
	S \circ f = f \circ S \;.
\]
The collection of all reversors and symmetries of a map form a group, called the \textit{reversing symmetry group} \cite{Lamb98}.
For example, when the force is periodic, $F(x+1) = F(x)$, then \eqref{eq:genStd} commutes with the integer rotation:\footnote
%%%%
{
	Here we are really considering the lift of the map to the plane, by undoing 
	the mod 1 operation. It is convenient to do this to count the number of 
	rotations of an orbit. Indeed, as was suggested to me by Robert MacKay, 
	\Std~uses a structure, \mfont{ifnumber}, for
	$x$ containing a \mfont{long} and \mfont{double} for the integer and fractional parts.
}
\[
	R(x,y) = (x+1,y) \;.
\]%%%%
It is easy to see that the composition of any symmetry with a reversor is itself a reversor. Thus the maps
\begin{align*}
	S_3(x,y) &\equiv S_1 \circ R(x,y) = (-1-x, y+F(x)) \;,\\
	S_4(x,y) &\equiv S_2 \circ R(x,y) = (y-x-1, y) \;,
\end{align*}
are also reversors; they give another decomposition, $f = S_4 \circ S_3$ The fixed sets of these reversors are
\begin{align*}
	\Fix{S_3} &= \{(-\frac12, y)\} \;, \\
	\Fix{S_4} &= \{(x,y): y = 2x + 1 \} \;,
\end{align*}
since when the force is both periodic and odd, then $F(\frac12) = F(-\frac12) = 0$.

The map $f$ is itself a symmetry, since it commutes with itself. This implies generally that the transformations $f^n \circ S$, where $S$ is a reversor, are also reversors. The set of reversors generated by $\{S_1, R, f\}$ form a \textit{family}.

You can plot the fixed sets of this family in \Std~with the command \menu{Find}{Symmetry Lines...}, \cmd-L. This command will draw the four fixed sets $\Fix{S_i}$, and by clicking the \mfont{Iterate} button you can iterate generate the fixed sets $\Fix{f^{2k} \circ S_i}$: the image of a fixed set is the fixed set of another member of the family:
\[
	f(\Fix{S}) = \{ z: f^{-1}(z) = S(f^{-1}(z)) \} = \Fix{ f^2 \circ S} \;.
\]
Inverse images of the symmetry lines can be found by selecting
\menu{Change}{Inverse Iteration}. Several forward and backward iterates of the symmetry lines for \eqref{eq:stdmap} are shown in \Fig{symmetry}.

The most interesting aspects of the resulting picture are the many intersections of the symmetry sets. It is not hard to see that fixed sets of reversors in a given family intersect at symmetric periodic orbits; for example, if $z \in \Fix{S} \cap \Fix{f^{2k} \circ S}$, then
\[
	z = S(z) = f^{2k}S(z)  = S(f^{2k} (S(z))) = f^{-2k}(z) \;,
\]
so $z$ has period $2k$. Several orbits near such symmetric periodic orbits are shown in \Fig{symmetry}.

%%%%%
\InsertFig{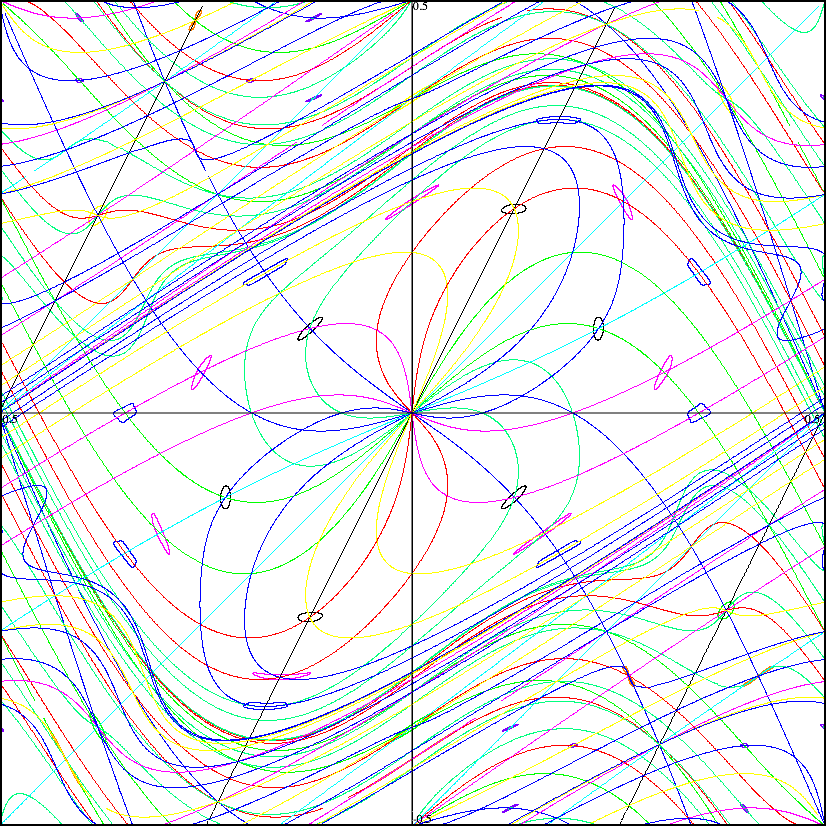}{Four forward and four backward iterates of the symmetry lines for \eqref{eq:stdmap} at $k = 1.3$ and several orbits near some elliptic, symmetric orbits.}{symmetry}{3in}
%%%%%
The generalized standard map also has the inversion
\[
	I(x,y) = (-x,-y)
\]
as a symmetry when $F$ is odd. This symmetry can be used in \Std~ to ``mod" the dynamics to the upper half-plane, by selecting \menu{Change}{Clipping} $\to$ \mfont{ y > 0}; in this case points with $y < 0$, are reflected to the positive half-plane using $I$. The inversion symmetry 
gives rise to two additional reversors
\begin{align*}
	S_5(x,y) &= S_1 \circ I(x,y) = (x, -y -F(x)) \;,\\
	S_6(x,y) &= S_2 \circ I(x,y) = (x-y, -y) \;.
\end{align*}
This ``extra'' symmetry gives rise to some otherwise surprising bifurcations of the librating periodic orbits \cite{MacKay84b} are it has also been profitably used in the study of nontwist maps \cite{Wurm05}.
%\begin{align*}
%	\Fix{S_5} &= \{(x,y) : y = -\frac12 F(x))\} \\
%	\Fix{S_6} &= \{(x,0)\}
%\end{align*}

Finally, the generalized standard map also has a vertical rotation symmetry, $V(x,y) = (x,y+1)$, in the sense that
\beq{vertical}
	f \circ V =  R\circ V\circ f \;.
\eeq
This symmetry can be exploited in \Std~to ``mod" orbits to the range $-\frac12 \le y < \frac12$ or in conjunction with the inversion $I$ to $ 0 \le y < \frac12$. Both of these settings are available on \menu{Change}{Clipping}.

Since the fixed sets of the orientation-reversing involutions are one-dimensional, it is  easier to search for symmetric periodic orbits than general orbits. \Std~exploits this by using a one-dimensional secant method to find symmetric orbits when you select \menu{Find}{Periodic Orbit...}. To find an $S$-symmetric period-$q$ orbit one could look for fixed point of $f^q$ that lies on $\Fix{S}$. However, it is a bit better to halve the work, as follows. Suppose $z$ is a $(p,q)$, orbit, i.e., after $q$ iterates, the orbit rotates $p$ times so $f^q(z) = R^p(z)$. If, in addition $z \in \Fix{S}$ and $q$ is even then
\[
	f^{q/2}(z) = f^{q/2}(S(z)) = S(f^{-q/2}(z)) = S(R^{-p}(f^{q/2}(z)))
\]
which implies that $z_k \in \Fix{S\circ R^{-p}}$, where $k = q/2$. Thus to search for such an orbit we need to iterate only $k$ times starting on $\Fix{S}$ and ending on $\Fix{S \circ R^{-p}}$. When $q$ is even $p$ should be odd, since $(p,q)$ should have no common factors. Moreover, since $S_1$ flips the sign of $x$,
$\Fix{S_1 \circ R^{-p}} = R^{(p+1)/2} \Fix{S_1 \circ R}$, that is, the fixed set $\Fix{S_3}$ rotated $(p+1)/2$ times.  The full list of possible symmetric orbits depending upon whether $p$ and $q$ are odd or even is given in \Tbl{symmetry}.  For each $(p,q)$ there are two different symmetric orbits, and they each have points on two of the symmetry sets.

\begin{table}
\centering
\begin{tabular}{c|c|c|c}
 $(p,q)$  &	$z_0$		&	$z_k$ & $(l,k)$ \\
\hline
(odd,even)	 &$\Fix{S_1}$	&	$\Fix{S_3}$ & 	$(\frac{p+1}{2}, \frac{q}{2}$) \\
	&		  $\Fix{S_2}$	&	$\Fix{S_4}$ &	$(\frac{p+1}{2}, \frac{q}{2}$) \\	 
	&		  $\Fix{S_3}$ 	&	$\Fix{S_1}$ & 	$(\frac{p-1}{2}, \frac{q}{2}$) \\
	&		  $\Fix{S_4}$ 	&	$\Fix{S_2}$ &	$(\frac{p-1}{2}, \frac{q}{2}$) \\
\hline
(even,odd)	&$\Fix{S_1}$		&	$\Fix{S_2}$ &	$(\frac{p}{2}, \frac{q+1}{2})$ \\
	&		 $\Fix{S_2}$		&	$\Fix{S_1}$ &	$(\frac{p}{2}, \frac{q-1}{2})$ \\
	&		 $\Fix{S_3}$		&	$\Fix{S_4}$ &	$(\frac{p}{2}, \frac{q+1}{2})$ \\	
	&		 $\Fix{S_4}$		&	$\Fix{S_3}$ &	$(\frac{p}{2}, \frac{q-1}{2})$ \\
\hline
(odd,odd)	&$\Fix{S_1}$		&	$\Fix{S_4}$ &	$(\frac{p+1}{2}, \frac{q+1}{2})$ \\
	&		 $\Fix{S_2}$		&	$\Fix{S_3}$ &	$(\frac{p+1}{2}, \frac{q-1}{2})$ \\
	&		 $\Fix{S_3}$		&	$\Fix{S_2}$ &	$(\frac{p-1}{2}, \frac{q+1}{2})$ \\
	&		 $\Fix{S_4}$		&	$\Fix{S_1}$ &	$(\frac{p-1}{2}, \frac{q-1}{2})$ \\
\end{tabular}
\caption{Symmetry lines containing $(p,q)$, symmetric periodic orbits for a reversible map with reversor $S_1$ and discrete rotation symmetry $R$. Here $S_2 = f \circ S_1$, $S_3, = S_1 \circ R$ and $S_4 = S_2 \circ R$. The initial point $z_0$  on the symmetry line shown maps to the point $z_k$ on the second symmetry line after $l$ rotations. }
\label{tbl:symmetry}
\end{table}

When you select \menu{Find}{Periodic Orbit...}, the dialog box has a \mfont{symmetry} pop-up menu where you can select the beginning symmetry line for the $(p,q)$ orbit; this menu has six entries. The last four entries allow you to select one of the four fixed sets for the initial point of the orbit. A number of symmetric orbits are shown in \Fig{SymmetricOrbits}. As \Tbl{symmetry} shows, one of the symmetric orbits always has a point on $\Fix{S_1}$; it can be shown that this orbit has positive residue when $k >0$: $\Fix{S_1}$ is called the \textit{dominant} symmetry line for the standard map.\footnote
%%%
{
	When $k < 0$ the dominant line for \eqref{eq:stdmap} is $\Fix{S_3}$. 
	Other reversible maps, like the H\'enon map, \cmd-2, also appear to have 
	dominant symmetry lines, though as far as I know this has not been proved.
}
%%%
All of the elliptic islands for the rotational orbits line-up on this fixed set. The second orbit is always hyperbolic, and either has a point on $\Fix{S_1 \circ R} = \Fix{S_3}$ when $q$ is odd, or $\Fix{f \circ S_1} = \Fix{S_2}$ when $q$ is even. 

The first two entries in the \mfont{symmetry} menu,  \mfont{Minimizing} and \mfont{Minimax} refer to the action-minimizing and minimax orbits of Aubry-Mather theory, see \cite{Meiss92} for a review.  The minimizing orbit is always hyperbolic \cite{MacKay83} and corresponds to the second orbit above. The minimax orbit has positive residue and, when $k$ is positive and small enough, it is elliptic. It has a point on the dominant symmetry line.

%%%%%
\InsertFig{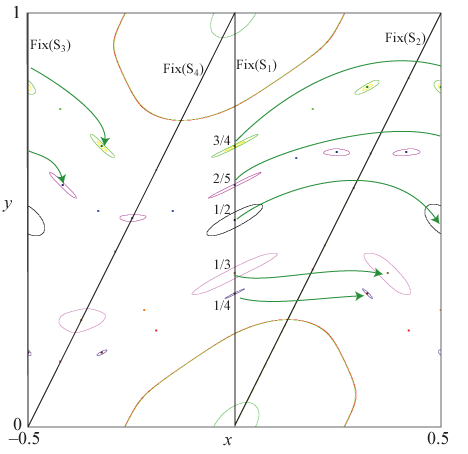}{Symmetry lines and some symmetric orbits for the standard map with $k = 1.0$.}{SymmetricOrbits}{4in}
%%%%%

%%%%%%%%%%%%
%%%%% Greene
%%%%%%%%%%%%
\section{The Critical Golden Circle}\label{sec:greene}

The $(0,1)$ resonance surrounding the elliptic fixed point is enclosed by a connected chaotic region generated by the unstable manifold of the $(0,1)$ saddle orbit.
Chirikov noticed that this chaotic region appears to be bounded when $|k| \lesssim 1$ and unbounded for larger values of $|k|$ \cite{Chirikov79}. One way to see this is to look for \textit{climbing} orbits, that is orbits that move from $y = a$ to $y = b$ for $a < b$. 
\Std~provides a convenient way to do this experiment: select \menu{Find}{Transit Time...}. The program then asks you to drag the mouse over a rectangle, $
\mathcal{R}_i$, in which the initial conditions will be selected---for example, choose a rectangle near the fixed point $(0.5,0)$. A dialog window will open that shows the boundaries of the rectangle that you selected and also has a field labeled $N$, which will be the maximum length of any attempted transiting orbits; by default, $N = 10^5$. Clicking on the \mfont{OK} button will accept these values, and you will be asked to drag over a rectangle $\mathcal{R}_f$ defining the final region. Choose a small rectangle near the point on the $(1,2)$ (minimizing) saddle orbit.\footnote
%%%
{
	Near, e.g., $(0.306, 0.612)$  when $k = 1.5$.
}
%%%
After you accept the values for the final rectangle, \Std~will iterate randomly selected initial conditions in $\mathcal{R}_i$ until they happen to fall in $\mathcal{R}_f$, or until the number of iterates reaches $N$. 

When you click on the \mfont{Stop} button, \Std~will provide you with a binned list of the transit times. While \Std~does not make a plot of this histogram (file a bug report!) it is easy to copy the data from the text window (Select it and type \cmd-C). It can be imported into another program such as Maple\texttrademark~or Matlab\texttrademark~to make the plot, see \Fig{histogram}. The average transit time for $k = 1.5$  is $t_{trans} = 501$ steps, at  $k = 1.1$ is has grown to $t_{trans} = 3.7\times10^4$, and when $k = 1.0$ it has become $3.9\times 10^6$. In each case the transit-time distribution looks roughly exponential, but experience shows that the long-time distribution will decay only algebraically with time \cite{Karney83, Hanson85, Meiss92}. Since the average transit time grows rapidly as $k$ decreases towards one,\footnote
{
   The time grows as $t_{trans} \sim 75 (k-k_{cr})^{-3.01}$ \cite{MacKay84}.
}
this technique is not a good one in which to determine the critical parameter value, $k_{cr}$, at which $t_{trans}$ first becomes $\infty$.

%%%%%
\InsertFig{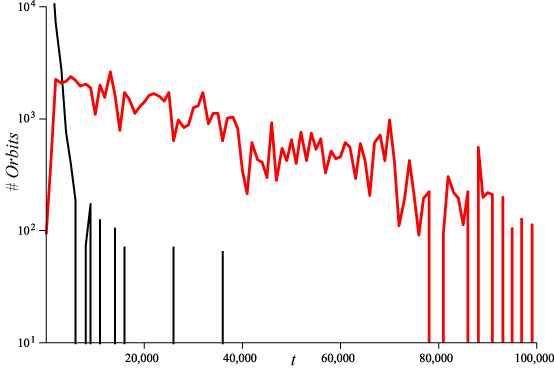}{Histogram of transit times for orbits beginning near the $(0,1)$ saddle and ending near the $(1,2)$ saddle for $k = 1.5$ (black) and $k = 1.1$ (red).}{histogram}{3in}
%%%%%

A much better way was discovered by John Greene \cite{Greene79, Greene80, MacKay93}. This is to look for the ``last" rotational invariant circle that separates the rectangles $\mathcal{R}_i$ and $\mathcal{R}_f$. Indeed, Birkhoff showed that if there is no orbit that reaches $\mathcal{R}_f$ then there must indeed be such an invariant circle \cite{Meiss92}, and Mather provided a variational construction of climbing orbits \cite{Mather91}.

Greene's insight was to look at the stability properties of periodic orbits that approximate an invariant circle. His hypothesis was that when these nearby orbits were ``stable", the circle should exist, and when they were unstable, it should be destroyed. A persistent invariant circle has an irrational rotation number $\omega$ and, as we learn from KAM theory, $\omega$ should satisfy a Diophantine condition \cite{Meiss92}. Greene used the rotation number to select a family of periodic orbits defined as the convergents of the continued fraction expansion of $\omega$. That is, suppose that
\[
	\omega = a_0 + \frac{1}{a_1 + \frac{1}{a_2 + \ldots}} = [a_0,a_1,a_2,\ldots] \;,
\]
where $a_i \in \N$ are the continued fraction elements. Then 
\[
	\omega_i = \frac{p_i}{q_i} = [a_0, a_1, \ldots, a_i]
\]
is the $i$th convergent to $\omega$. Greene computed the residues $R_i$ of these the $(p_i,q_i)$ orbits, and observed that when $k < k_{cr}(\omega)$, $R_i \to 0$, and the invariant circle exists. When $k > k_{cr}(\omega)$, $R_i \to \infty$ and the circle is destroyed. At the critical parameter, $R_i$ remains bounded and the invariant circle exists but appears to be nonsmooth.

At the time, Greene's observation seemed quite preposterous to mathematicians familiar with the number theoretic delicacies of KAM theory. Nevertheless, Greene's criterion works extremely well and many aspects of his observations were subsequently proven \cite{MacKay92, Delshams00}.

Greene provided strong numerical evidence that the last rotational invariant circle for the standard map has $\omega = \gamma$, where $\gamma = \frac{1+\sqrt{5}}{2}$ is the golden mean. The symmetries of \eqref{eq:stdmap}, discussed in \Sec{reversible}, imply that circles with rotation numbers $\pm \omega + m$ for any $m \in \Z$ are equivalent. This means, for example that the circles $\gamma^{-1} = \gamma-1$ and $\gamma^{-2} = 2-\gamma$ are also destroyed at $k_{cr}(\gamma)$.

This golden mean is distinguished by its continued fraction expansion $\gamma = [1,1,1,1,\ldots]$; it is, in the Diophantine sense, the most irrational number.  Percival called numbers whose continued fractions have an infinite tail of $1$'s, \textit{noble} numbers, and it appears that the noble invariant circles are locally the most robust circles \cite{Schmidt82,Buric90, MacKay92b}.

To find a sequence of convergent periodic orbits in \Std, use the 
\menu{Find}{Farey Path...} command. The 
dialog window that opens will ask for a pair of numbers $(p_0,q_0)$ and $(p_1,q_1)$ that will form the base of a \textit{Farey tree}; these numbers must be neighbors in the sense that
\[
	p_1 q_0 - p_0 q_1 = 1 \;.
\]
The default is to choose $(0,1)$ and $(1,1)$, which will allow you to construct any number $0 < \omega < 1$.  We start with the triplet $\{\omega_L,\omega_C, \omega_R\}$ where
$\omega_L = \frac{p_0}{q_0}$ is the lower neighbor, $\omega_R = \frac{p_1}{q_1}$ is the upper neighbor and 
\[
	\omega_{C} = \omega_L \oplus \omega_R \equiv \frac{p_0 + p_1}{q_0+q_1}
\]
where $\oplus$ is the \textit{Farey sum} operation. 

The Farey path of a number $\omega \in (\omega_L, \omega_R)$ is the unique sequence of \textit{left} and \textit{right} turns that lead to $\omega$ using the recursive algorithm:

\begin{enumerate}
	\item If $\omega = \omega_C$ stop.
	\item If $\omega \in (\omega_L, \omega_C)$ then the step is $L$, 
the next rotation number is
$
	\omega_C' = \omega_L \oplus \omega_C \;,
$
and the triplet becomes $\{\omega_L, \omega_C', \omega_C\}$.
	\item If $\omega \in (\omega_C,\omega_R)$ then the step is $R$,  
$
	\omega_C' = \omega_C \oplus \omega_R \;,
$ 
and the triplet is $\{\omega_C,\omega_C', \omega_R\}$. 
	\item Repeat.
\end{enumerate}

In each case the outer pair of the triplet are Farey neighbors. The path truncates if and only if $\omega$ is rational. It is not hard to see that noble irrationals have a Farey path that eventually oscillates $ \ldots LRLRLRLR \ldots $.

Typing a sequence of $L$'s and $R$'s (or $l$'s and $r$'s or $0$'s and $1$'s) in the text box for the \mfont{Path}, instructs \Std~to find each of the orbits with rotation numbers on that path. If you turn on the computation of the residue, \menu{Change}{Compute Residue}, then the Residue is calculated as well. The default settings, $\omega_L = \frac01$, $\omega_R = \frac12$ and the path of alternating $LR$'s gives a sequence that converges to $\omega = \gamma^{-2}$; the results are shown in \Tbl{critical}.  Here we have computed the minimax periodic orbits that sit on the line $\Fix{S_1}$, recall \eqref{eq:dominant}.

%%%%%%%%%%%%%%%%%%%%%%%%
\begin{table}
\centering
\begin{tabular}{c|l|l|l|l}
$(p,q)$ & $y_0 (k=k_{cr})$ & $R(k=0.971)$ & $R (k_{cr})$ & $R (k = 0.972)$ \\
\hline
(0,1)	 & 	0.0	 					&0.2425 			& 	0.24290885	& 	 0.243\\ 	 
(1,1)	 & 	1.0		 				&0.2425 			& 	0.24290885	& 	 0.243\\ 	 	 
(1,2)	 & 	0.5		 				&0.235225		& 	0.23601884	& 	 0.236196\\ 	 	 
(1,3)	 & 	0.370744011711781		&0.25936027 		& 	0.26068032	& 	 0.26097522\\ 	 
(2,5)	 & 	0.415872871308091		&0.23996369 		& 	0.24201467 	& 	 0.24247383\\	 
(3,8)	 & 	0.401519832446222		&0.25204405 		& 	0.25552012	& 	 0.25630081 \\	 
(5,13)	 & 	0.406236169619200		&0.24114743		& 	0.24660815	& 	 0.24784114\\ 	 
(8,21)	 & 	0.404701675915617		&0.24328205 		& 	0.25229896 	& 	 0.25435250 \\	 
(13,34)	 & 	0.405202192064269		&0.23440617 		& 	0.24872040	& 	 0.25202636 \\	 
(21,55)	 & 	0.405038969799706		&0.22786224 		& 	0.25093749	& 	 0.25638918 \\ 	 
(34,89)	 & 	0.405092179098598		&0.21330551 		& 	0.24956827 	& 	 0.25845593\\	 
(55,144)	 & 	0.405074826384691		&0.19394915 		& 	0.25040872 	& 	 0.26508896\\ 
(89,233)	 & 	0.405080483747179		&0.16490212 		& 	0.24989322 	& 	 0.27418217\\ 
(144,377)	 & 	0.405078638932727	&0.12726102 		& 	0.25020866 	& 	 0.29100469\\ 
(233,610)	 & 	0.405079240426684	&0.08331189 		& 	0.25001571 	& 	 0.31974102\\ 
(377,987)	 & 	0.405079044294881	&0.04197067 		& 	0.25013371 	& 	 0.37342358\\ 
(610,1597)	 & 	0.405079108244936	&0.01380301 		& 	0.25006166 	& 	 0.48046976\\ 
(987,2584)	 & 	0.405079087392894	&0.00228248 		& 	0.25010655 	& 	 0.72560041\\ 
(1597,4181)	 & 	0.405079094191927	&1.24150225E-4	& 	0.25007802 	& 	 1.42403113\\ 
(2584,6765)	 & 	0.405079091975001	&5.08240191E-7	& 	0.25007892 	& 	 4.31242653\\ 
(4181,10946) & 	0.405079092697857	&6.96835923E-6	& 	0.25014343 	& 	 26.8778283\\
(6765,17711) &  0.405086090209096	&5.35564322E-6	&	0.24978752	&	 553.873723\\
(10946,28657) & 0.405086089756233	&-4.87898942E-5	&	0.24703952	& 	 1.12588415E+5 \\
\end{tabular}
\caption{Residues of the convergents of the minimax periodic orbits, with $x_0 = 0$ on the Farey path for the rotation number $\gamma^{-2}$ for three values of $k$, including $k_{cr} = 0.97163540631$.}
\label{tbl:critical}
\end{table}
%%%%%%%%%%%%%%%%%%% 

The values of $y_i$ for $k_{cr}$ appear to converge to $0.405086089$ as $i \to \infty$. This convergence is geometrical and can be understood by a renormalization argument \cite{MacKay92}. The residues for $k = 0.971$ do appear to converge to $0$, though the last value is not believable; careful computations show that the minimax orbits always have positive residues. Similarly, the residues for $k = 0.972$ grow, and those for $k = k_{cr}$ appear to be bounded, and moreover to converge to $0.25009$ \cite{MacKay92}: the last couple of values in \Tbl{critical} again have numerical convergence problems. 

When $k > k_{cr}$ the invariant circle no longer exists, but the beautiful theory of Aubry and Mather \cite{Aubry83, Mather82} implies that there still is an invariant set with the same rotation number; this set is a Cantor set, and Percival called it a \emph{cantorus}. Cantori can seen in \Std~using the Farey path command as above and setting $k > k_{cr}$. Several examples are shown for the rotation number $\gamma^{-2}$ in \Fig{cantori}. The gaps in the Cantor set appear to all be generated by images of a single gap. The largest gap is centered on the dominant symmetry line, $x = 0$, where all the elliptic orbits line up. This perhaps visualizes in the most precise way Chirikov's concept of resonance overlap \cite{Chirikov79}: the islands due to the different periodic orbits that approximate the irrational rotation number squeeze the invariant circle most strongly where they line up.

It is also easy to find cantori for other rotation numbers by changing the path in the Farey sequence. Since noble numbers are most robust, the cantori with the smallest gaps will be found by selecting paths that have a short head sequence, followed by an alternating LR-tail.

%%%%%
\InsertFig{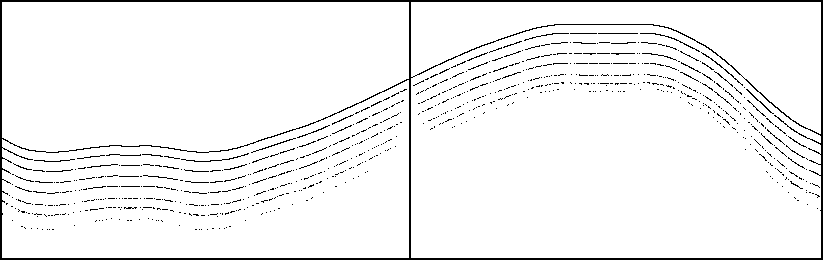}{Golden mean invariant circle and approximations to the cantori for the standard map at $k = k_{cr}$, $0.974$, $0.977$, $0.980$, $0.985$, $0.990$, $1.0$ and $1.05$. The cantori have been shifted vertically to avoid overlap.}{cantori}{6in}
%%%%%

When an invariant circle is destroyed, the resulting cantorus has a nonzero flux of crossing orbits. This flux can be computed by taking the difference between the action of the minimax and minimizing orbits that approximate the invariant Cantor set, see \cite{Meiss92} for a discussion. The action of orbits can be computed in \Std~by selecting \menu{Change}{Compute Residue \& Action}.

%%%%%%%%%%%%%%%%%
%% Other systems
%%%%%%%%%%%%%%%%%
\section{Exploring other Maps}\label{sec:others}

\Std~has thirteen reversible, area-preserving maps in the \mfont{Mapping} menu. If your favorite system is not included, do let me know and I will consider adding it.\footnote
{
	It would be best, of course, for me to add a parser to \Std~so that
	arbitrary maps could be iterated. However, much of the code depends upon
	the reversibility assumption, so this would not be easy.
}
Several representative phase portraits are shown in \Fig{otherMaps}. The last two dynamical systems in the \mfont{Mapping} menu are actually Poincar\'e sections of Hamiltonian flows. For example, the \textit{two-wave} Hamiltonian is
\[
	H(x,y,t) = \frac12 y^2 - \frac{1}{2\pi^2} \left [a \cos(2\pi x) + b \cos(2\pi(kx-t))\right]
\]
where $k \in \Z$.
This system has one and a half degrees of freedom, and as such is most properly described in a three-dimensional phase space $(x,y,t) \in \T \times \R \times \T$. The Poincar\'e map
$(x,y) \to (x',y')$ is the time-one map constructed by integrating the equations of motion for an initial condition $(x(0), y(0),0)$ to $t = 1$, and plotting the resulting values $(x(1),y(1))$. In \Std, this integration is done with a symplectic splitting method that is second-order accurate with a time step of $0.1$. It gives results that are qualitatively good, but should not be trusted for quantitative accuracy. I used it to generate the figures in \S9.16 of \cite{Meiss07a}.

%%%%%
\InsertFigFour{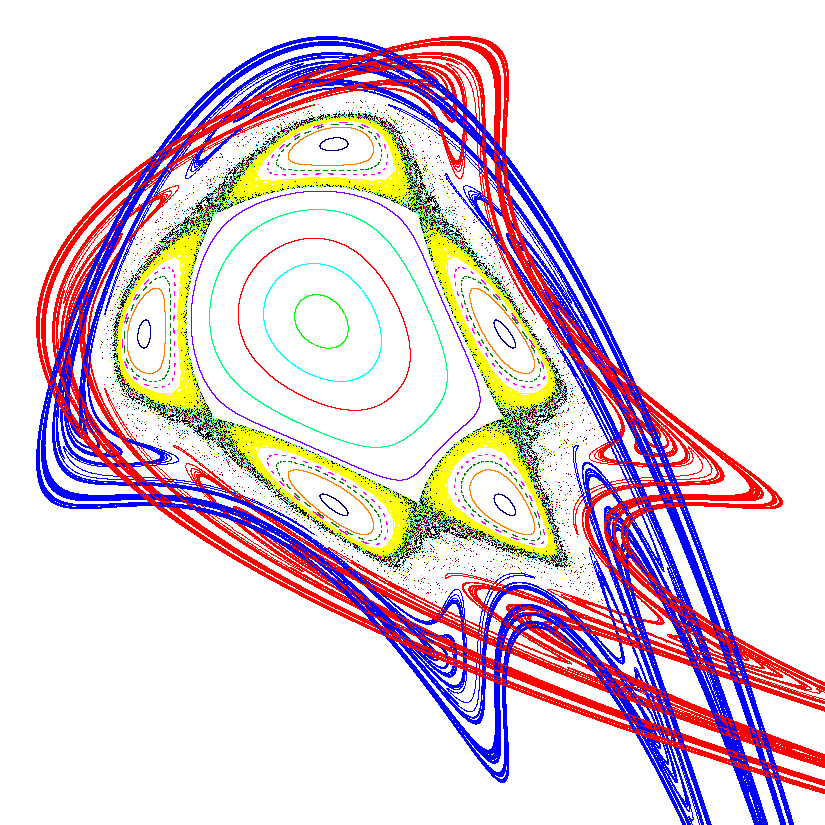}{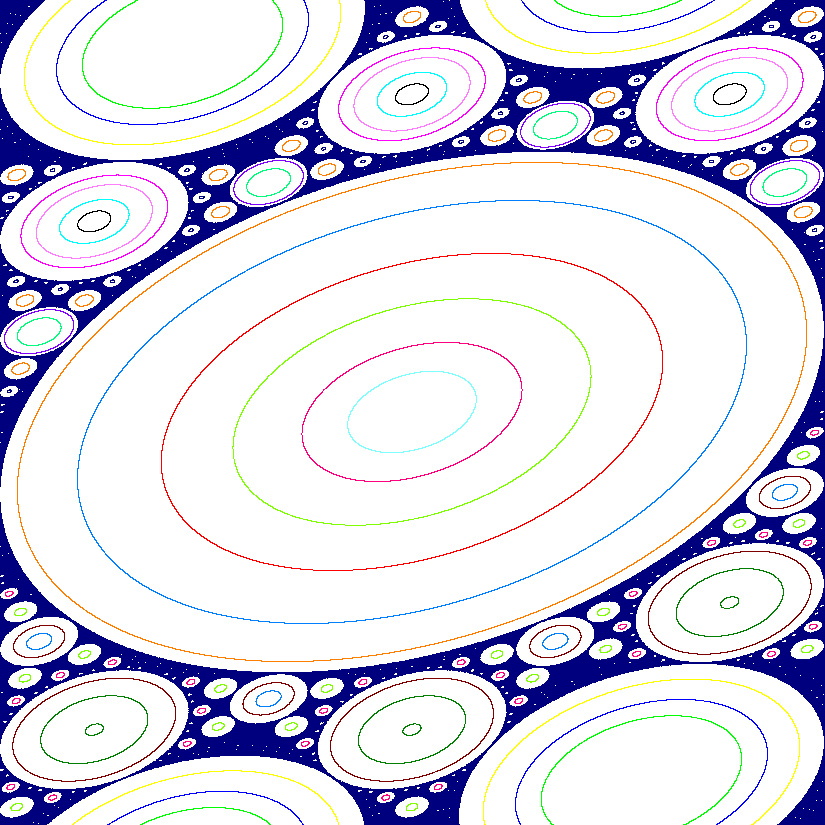}{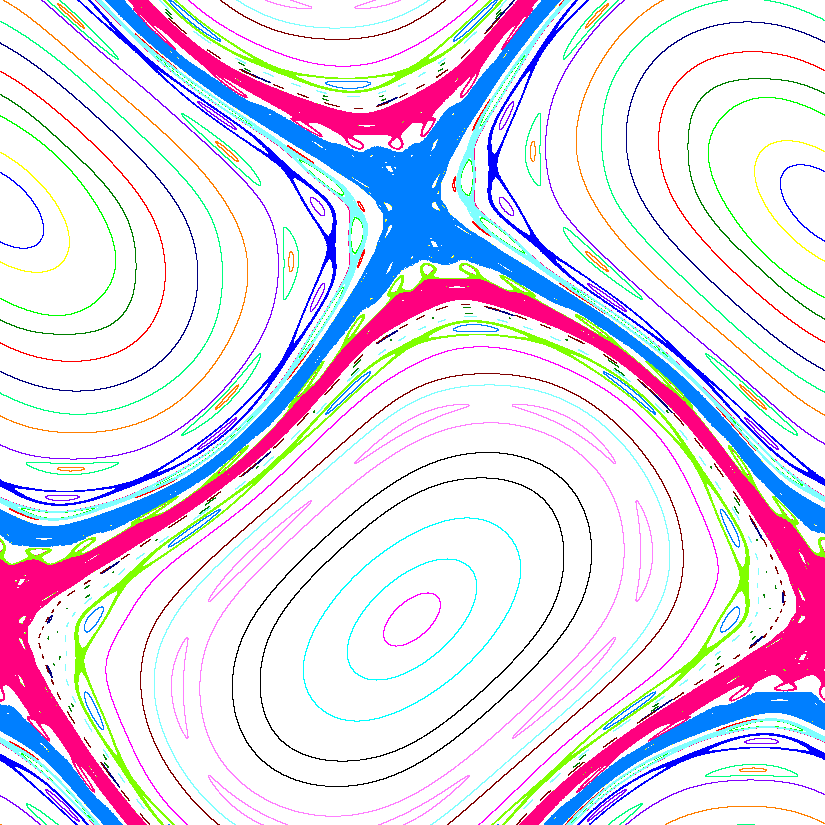}{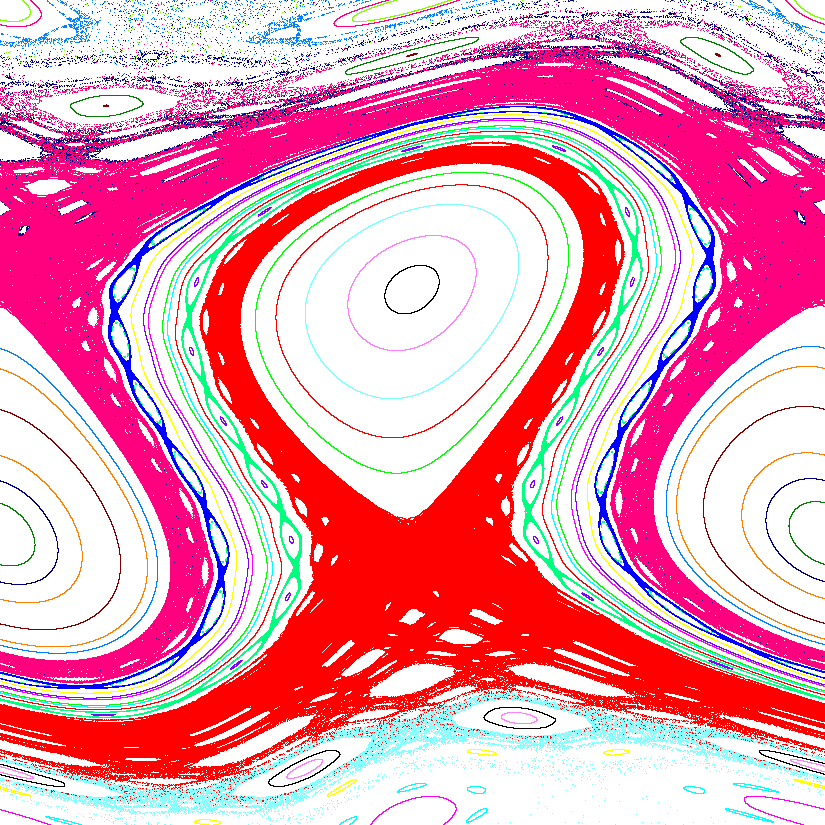}{Four representative phase portraits generated by \Std. Which of the thirteen systems do they represent?}{otherMaps}{3in}
%%%%%

\section{For Help}

\Std~has many other features that are not documented in this tutorial. In many cases the menu items should be self-explanatory; if not there are two help facilities. One is the \mfont{Help} menu which gives access to several files that briefly describe the features. A second is the \mfont{Help Tip} that pops-up if you let the mouse hover over a menu or button in the program. In some cases these tips have extended help that can be accessed by pressing the \cmd~key.

More discussion of the mathematics behind the standard map can be found in \cite{Meiss92, Meyer92, MacKay93, Easton98, Gole01}.

%%%%%%%%%%%
\bibliographystyle{unsrt}
\bibliography{stdmap}

\end{document}